**Title:** Mapping Whole Exome Sequencing to *In Vivo* Imaging with Stereotactic Localization and Deep Learning

**Authors:** Mahsa Servati, MSc[1,2], Courtney N Vaccaro, MSc[3], Emily E Diller, PhD[4], Renata Pellegrino Da Silva, PhD[3], Fernanda Mafra, PhD[5], Sha Cao, PhD[1], Katherine B. Stanley, BSc[1], Aaron Cohen-Gadol, MD, MSc, MBA[1], Jason G Parker, PhD[1,2]

**Affiliations:**
[1]School of Medicine, Indiana University, Indianapolis, IN, USA
[2]School of Health Sciences, Purdue University, West Lafayette, IN, USA
[3]Center for Applied Genomics, Children's Hospital of Philadelphia, Philadelphia, PA, USA
[4]Feinberg School of Medicine, Northwestern Medicine, Chicago, IL, USA
[5]10x Genomics, Pleasanton, CA, USA

**Corresponding Author's name and current institution:**
Jason G Parker, PhD
Radiology and Imaging Sciences, School of Medicine, Indiana University
950 W. Walnut St., R2 E159B
Indianapolis, IN 46202-5188
(317) 274-2072 tel
(317) 274-1067 fax

**Corresponding Author's Email:**
parkerjg@iu.edu

**Abstract**

*Objective:* Intra-tumoral heterogeneity (ITH) complicates the diagnosing and treating glioblastoma. Although multiparametric imaging enhances the characterization of ITH—by evaluating both spatial and functional variations in phenotype—significant limitations exist in assessing and visualizing cellular and molecular properties across space. These limitations are due to a lack of easily accessible, co-located pathology and genomic data. Stereotactic localization accurately maps biopsy results to their original locations in pre-surgical images; however, this technique is seldom employed as it can be difficult to implement in a surgical context given its specialized procedural requirements. This study presents a multi-faceted approach combining stereotactic biopsy with standard clinical open-craniotomy for sample collection, voxel-wise analysis of MR images, regression-based Generalized Additive Models (GAM), and whole-exome sequencing. This work aims to demonstrate the potential of machine learning algorithms to predict variations in cellular and molecular tumor characteristics.

*Methods:* This retrospective study enrolled ten treatment-naïve patients with radiologically confirmed glioma (5 WHO grade II, 5 WHO grade IV). Each patient underwent a multiparametric MR scan ($T1_W$, $T1_{W-CE}$, $T2_W$, $T2_W$-FLAIR, DWI) prior to surgery (27.9±34.0 days). During standard craniotomy procedure, at least 1 stereotactic biopsy was collected from each patient, with screenshots of the sample locations saved for spatial registration to pre-surgical MR data. Whole-exome sequencing was performed on flash-frozen tumor samples, prioritizing the signatures of five glioma-related genes: IDH1, TP53, EGFR, PIK3CA, and NF1. Regression was implemented with a generative additive model (GAM) using a univariate shape function for each predictor. Standard receiver operating characteristic (ROC) analyses were used to evaluate detection, with AUC (area under curve) calculated for each gene target and MR contrast combination.

*Results:* The mean AUC for the five gene targets and 31 MR contrast combinations was 0.75±0.11; individual AUCs were as high as 0.96 for both IDH1 and TP53 with $T2_W$-FLAIR and ADC and 0.99 for EGFR with $T2_W$ & ADC. An average AUC of 0.85 across the five mutations was achieved using the combination of $T1_W$, $T2_W$-FLAIR, and ADC.

*Conclusion:* These results suggest the possibility of predicting exome-wide mutation events from non-invasive, *in vivo* imaging by combining stereotactic localization of glioma samples and a semi-parametric deep learning method. This approach holds potential for refining targeted therapy by better addressing the genomic heterogeneity of glioma tumors.

## Introduction

Increasing mortality rates associated with brain tumors have highlighted a critical need for advancements in both diagnostic and therapeutic approaches.[1] The conventional diagnosis procedure involves pre-surgical imaging and one biopsy sample to assess cellular and molecular properties.[2] Such a procedure allows subsequent chemotherapy and radiation treatments to be optimized for patient-specific mutation profiles. However, a growing body of both basic and clinical evidence has demonstrated that the somatic and genomic composition of human brain tumors is not uniform across space and time.[3] This phenomenon, referred to as intra-tumoral heterogeneity (ITH), can compromise the accuracy of results obtained from conventional surgical biopsy, potentially rendering subsequent molecular characterizations incorrect. Though genomic instability is the primary driver of tumor heterogeneity,[4] clonal cell subpopulations show plasticity, shifting between cell states.[5] Their proliferation potential is influenced not only by their genetic composition but also by epigenetic factors like DNA methylation and changes in histone structures.[6] ITH can exist as variability in the gene, transcript, or protein levels of distinct cell subpopulations — macro-heterogeneity, Figure 1. B — or within the cells belonging to the same subpopulation — micro-heterogeneity,[4] Figure 1. C. Therefore, not all the present mutations and expression pathways in the tumor microenvironment (TME) will be identified in the pathology analysis of a conventionally collected single biopsy sample. To overcome this limitation, a non-invasive tool to assess cellular and molecular tissue characteristics across the entire tumor bed and TME is essential.

MRI (Magnetic Resonance Imaging) is the standard pre-surgical diagnostic imaging procedure for brain tumors, serving as a non-invasive multi-purpose diagnosis and treatment planning tool.[7] The adoption of MRI was initially driven by its superior contrast resolution in neuroimaging.[8] Now, beyond rendering detailed anatomical and functional insights, it plays a central role in obtaining detailed molecular and cellular characteristics of brain tumors.[7] Combining manifold MRI sequences, known as mpMR (multiparametric magnetic resonance), enables simultaneous assessment of anatomical, functional, and cellular information from the tumor in only one imaging session. A standard brain MRI, with and without a contrast, provides physiological and morphological information about the brain tumor, such as edema and necrosis. Integrating MR-derived tumor characteristics with genetic data from biopsy sample analyses using machine learning techniques introduces a promising approach to effectively address the ITH. The foundational hypothesis of this study suggests that, through such integration, a model can be constructed to navigate the heterogeneity challenge, mapping underlying somatic and genomic aberrations with MR imaging signatures.

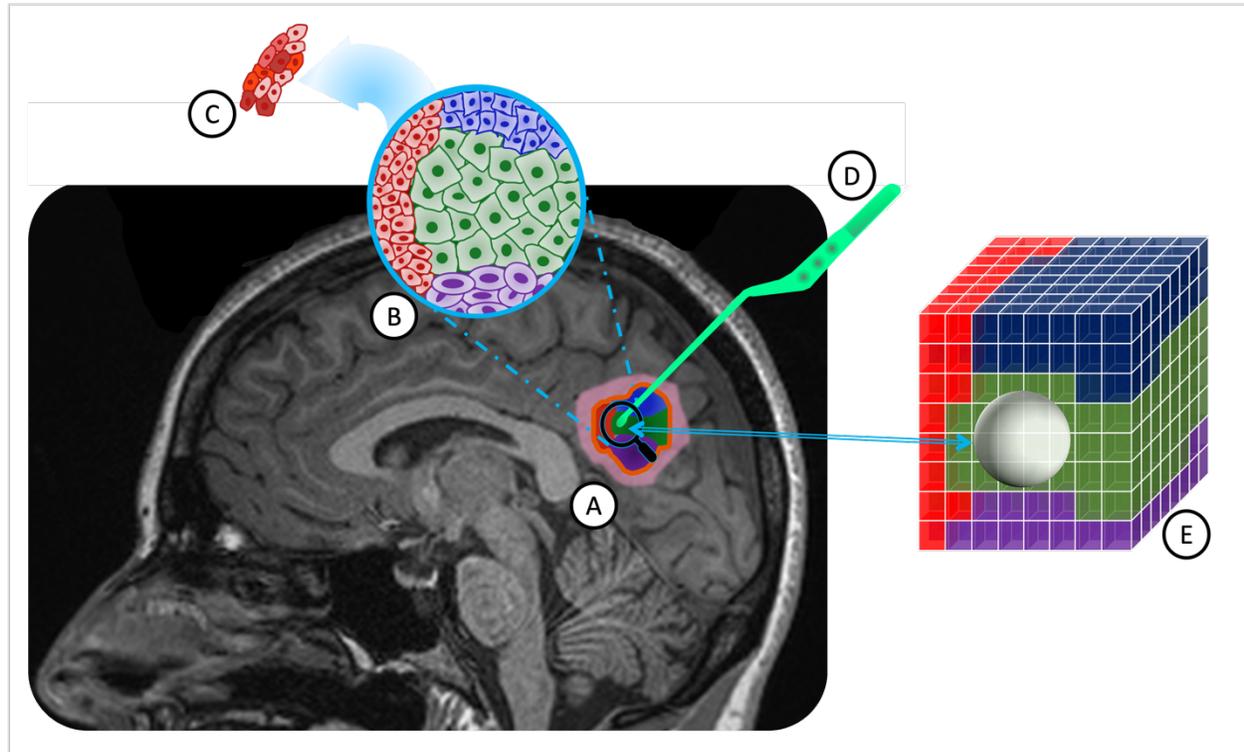

*Figure 1. ITH in a solid brain tumor and MRI-guided stereotactic biopsy, overlaid on a T1$_W$ MR image. **A.** A solid brain tumor outlined by a yellow border, comprising four distinct cell subpopulations (in red, blue, green, and purple) with associated inflammation in pink. **B.** Macro-heterogeneity within cell subpopulations, displaying reduced cytoplasm and enlarged nuclei in some cells. **C.** Micro-heterogeneity among tumor cells of the same subpopulations. **D.** The pituitary tweezer used for stereotactic biopsy. **E.** Tumor sample volume represented as a sphere, centered on the biopsied location, and including two cell populations (blue and green).*
*(ITH: intra-tumoral heterogeneity; MRI: magnetic resonance imaging; T1$_W$: T1-weighted; MR: magnetic resonance)*

In a prior retrospective study,[9] the potential to predict mutational heterogeneity by utilizing a multiparametric MR-based machine learning algorithm in conjunction with advanced geometric modeling and random field theory was demonstrated. Notably, this initial cohort included a broad spectrum of brain diagnoses, not solely gliomas, and the genetic data was sourced from standard clinical pathology techniques, such as H&E staining and immunohistochemistry, limiting its depth and scope. The purpose of this current study was to extend our previous work in a population of pure glioma subjects. Imaging data were collected from the routine mpMR imaging acquired as a standard of care for each patient. During surgical intervention, at least one stereotactic tumor biopsy sample was collected prospectively, with the research team meticulously documenting the resection coordinates based on pre-surgical MR images. Immediately after collection, these samples were flash-frozen in the surgical suite and prepared for subsequent whole exome sequencing (WES). Subsequently, a semi-parametric machine learning model was built using the MR signals as predictors and genetic data as outcomes. In forthcoming sections, we present detailed methodology offering a potential avenue for understanding the complexities of tumor heterogeneity.

## Methods

*Protocol approval*

The Indiana University Institutional Review Board (IRB) approved and monitored this study as required by the United States 20 CFR Part 431.[10] This study presented no more than minimal risk to patients and qualified for expedited IRB review under categories 2 and 5. All subjects had previous MR imaging, but only a subset of subjects had available, usable blood samples. Informed consent was waived for all subjects with existing blood samples, and informed consent was obtained for all patients requiring a blood draw. Given that the data collected included patient health information (PHI), appropriate measures were taken to protect subject privacy and confidentiality. All procedures, methods, and experiments performed in this study were carried out in accordance with relevant guidelines and regulations, including the Declaration of Helsinki and the HIPAA Privacy Rule. This study was not listed on CinicalTrials.gov, and no part of the dataset presented here has been used or published in the past.

*Study population*

Ten patients (mean age, 47.0 ± 17.7 years; age range, 25-71 years; 7 males and 3 females) with radiologically diagnosed primary glioma grade II to IV according to the current WHO criteria[11] (5 WHO grade IV, 5 WHO grade II) were included in this study. One patient had a recurring diffuse astrocytoma grade II; all other enrolled subjects were treatment naïve. All subjects enrolled were scheduled for surgical resection of their brain tumor. Each patient had a tumor of sufficient size to guarantee the acquisition of at least one biopsy sample during the surgical procedure. As a part of their SOC, all patients underwent a clinical multiparametric MR scan within 27.9 ± 34.0 days prior to surgery. Table 1 presents details of the patient cohort.

*Table 1. Patient demographics and cohort*

| | | | |
|---|---|---|---|
| **Total patients** | 10 | **Radiologic diagnosis** | 10 WHO grade II-IV glioma |
| **Male** | 7 | **Pathologic diagnosis** | 5 glioblastoma, 2 oligodendroglioma (grade II), 3 astrocytoma (grade II) |
| **Age (mean ± STDV)** | 47.0 ± 17.7 years | $\overline{\Delta_{SX-IMG}}$ **(mean ± STDV)** | 27.9 ± 34.0 days |
| **Age range** | 25-71 years | $R_{SX-IMG}$ | 0-96 days |
| **Oncologic status** | 10 primary | **Number of samples (mean ± STDV)** | 1.8 ± 0.4 |

$\overline{\Delta_{SX-IMG}}$ : The average days between MRI session and surgery for all patients.
$R_{SX-IMG}$ : The range of days between MRI session and surgery for all patients.
(STDV: standard deviation; WHO: world health organization)

*Biopsy sampling and analysis*

Biopsy specimens, averaging 1cm$^3$ in size, were acquired by the clinical neurosurgeon (ACG) before initiating complete tumor resection. With the patient under general anesthesia and their head stabilized in a three-point head holder, a craniotomy was performed, guided by a frameless stereotactic system. Upon tumor exposure, ACG selected the biopsy location based on areas highlighted by fluorescein fluorescence.[12] Prior to sample collection, the neurosurgeon positioned the pituitary forceps on the target site (Figure 1.D.). Concurrently, research staff in the OR captured a screenshot from the stereotactic software (Medtronic Synergy Cranial v2.2.7), ensuring accurate recording of resection coordinates on the pre-surgical MR images. This method is the least disruptive to the patient's surgery but is known to be highly operator-dependent. Misregistration errors between the tumor tissue locations in the pre-surgical MR and the placement of the forceps by the surgeon have been shown in previous studies to result in errors of 2.4±1.7mm.[13]

Immediately following collection, each biopsy sample was flash-frozen in the operating room using liquid nitrogen to maintain genetic integrity. These frozen samples were sent to the Children's Hospital of Philadelphia (CHOP) for genomic analysis. From each sample, a minimum of 500ng of DNA was extracted. The DNA underwent whole-exome sequencing (WES) by the Center for Applied Genomics (CAG) at CHOP, utilizing the Twist Human Core Exome Capture Kit from TWIST Bioscience and the Illumina NovaSeq 6000 platform. Subsequently, demultiplexing of the acquired data was performed using the Illumina DRAGEN Bio-IT Platform (v3.6.3). The generated FASTQs were aligned against the Homo sapiens (GRCh37.75) reference with the DRAGEN pipeline as well, which implemented the Smith-Waterman Alignment Scoring algorithm. The WES samples underwent germline variant calling for single-nucleotide variant (SNV) and insertion/deletion variations. Somatic variant calling was then performed using a tumor-only protocol within the DRAGEN pipeline which excluded any germline variants. Variant call format (VCF) files were filtered for variants within a list of ten genes of interest. These filtered VCFs were used as input, along with the Glioblastoma multiforme Human Phenotype Ontology (HPO) term, into an internal variant annotation and prioritization tool to determine most relevant variants to this phenotype. At this point the number of outcome variables (i.e. genomic mutations) was too large to provide useful statistical results. To decrease the outcome variables to a reasonable number we chose to focus the remainder of this study on mutational status of eight gene targets: PTEN, IDH1, TP53, EGFR, PIK3R1, PIK3CA, NF1, and RB1, all known to be characteristic of glioma.[14] Three gene targets - PTEN, PIK3R1, and RB1 - were excluded from further analyses due to significant imbalance (specifically, only one positive patient for each).

The MRI datasets, which were obtained prior to biopsy resection as a part of standard of care for each patient, were identified and downloaded from local imaging centers. Seven patients underwent imaging using a 1.5T MRI scanner (5 Siemens and 2 GE), while two were imaged with a 3.0T Siemens MRI scanner. Owing to the extensive duration of the MRI session, one patient's imaging was conducted on both a 1.5T Siemens and then a 3.0T Toshiba scanner over two consecutive days.

All scans utilized a head-neck coil. Each patient's multiparametric imaging session comprised standard anatomical sequences: $T1_W$, $T1_{W-CE}$ (using a Gadolinium-based contrast agent), $T2_W$, and $T2_W$-FLAIR (fluid attenuated inversion recovery).[15,16] Additionally, a diffusion weighted imaging (DWI) sequence was employed to produce apparent diffusion coefficient (ADC) maps, offering both morphological and functional information. Due to clinical protocol variations, two patients were imaged using the $T2_{W-CE}$ sequence in place of $T2_W$.

*Image analysis*

Raw image data from $T1_W$, $T2_W$ (or $T2_{W-CE}$), $T2_W$-FLAIR, and ADC maps were co-registered to the $T1_{W-CE}$ frame-of-reference using the FMRIB Linear Image Registration Tool (FLIRT)[17,18] for each patient. A 12-degree-of-freedom (DOF) cross-correlation objective function was employed for $T1_W$ and $T2_W$-FLAIR registration,[18] while a 12-DOF mutual information objective function for $T2_W$ and DWI (only B0) registration.[19] The ADC map for each patient was then brought to alignment with their $T1_{W-CE}$ frame-of-reference using the affine transformation matrix estimated for the DWI B0 images.

Utilizing an automated white matter extraction tool,[20-22] all images from the five contrasts were normalized to the mean signal of an uninvolved Normal Appearing White Matter (NAWM) region on the $T1_{W-CE}$ image. The volume of tissue noted in the pathology report was used to draw an equal volume sphere on the $T1_{W-CE}$ image, centered on the location where the stereotactic biopsy was marked on three-plane neuro-navigation MR plans during surgery. This method ensured that the feature matrix and subsequent machine learning model were exclusively limited to the image voxels associated with the resected tissue, as shown in Figure 1. E.

*Statistical analysis*

Regression was implemented using the GAM (generalized additive model), a semi-parametric ensemble machine learning technique. The model was fitted to our dataset using the 'fitrgam' function available in the Statistics and Machine Learning Toolbox of MATLAB R2022a (MathWorks, Natick, MA, USA).[23] For each predictor ($x_i$), 'fitrgam' uses a univariate shape function ($f_i(x_i)$), which is typically a boosted tree representing a linear term for the predictor. Figure 2 provides an architectural visualization of the gradient boosting in GAM as a sequential ensemble. This algorithm incorporates weak learners, typically presented by decision stumps, with each successive tree aimed at correcting the errors of its predecessor. Following the addition of each tree, errors are evaluated via a loss function (L), with the ensemble adjusted to minimize mean-squared error. During its iterations, GAM identifies a learning rate (η) to reduce the deviance (D) for every observed response ($y_i$).

For each subject, distinct training and test feature matrices were formulated. These matrices comprised mpMR signatures as independent variables and binary class indicators representing the mutation status of target genes as dependent variables. A total of 1550 machine learning experiments were carried out, spanning 31 distinct combinations of 5 MR contrasts. These were evaluated against 5 gene targets for all 10 patients, resulting in 31x5x10 experiments. For each iteration, the training feature matrix was constructed using both imaging and genetic data from nine patients, reserving the tenth for the test feature matrix. This leave-one-patient-out methodology[24] ensured that every patient, across all 10 individuals, was singularly used as the test subject once, guaranteeing no overlap between training and testing data in any iteration. The 31 contrast combinations were categorized into five groups based on the number of contrasts: single, double, triple, quadruple, and quintuple contrasts. This categorization facilitated an analysis of the impact of combined predictors on the model's predictive efficiency. Furthermore, standard receiver operating characteristic (ROC) analyses were applied to every combination for each genetic target to determine the area under the curve (AUC) and accuracy (ACC) values, considering a threshold for FPR (False Positive Rate) of 0.2.

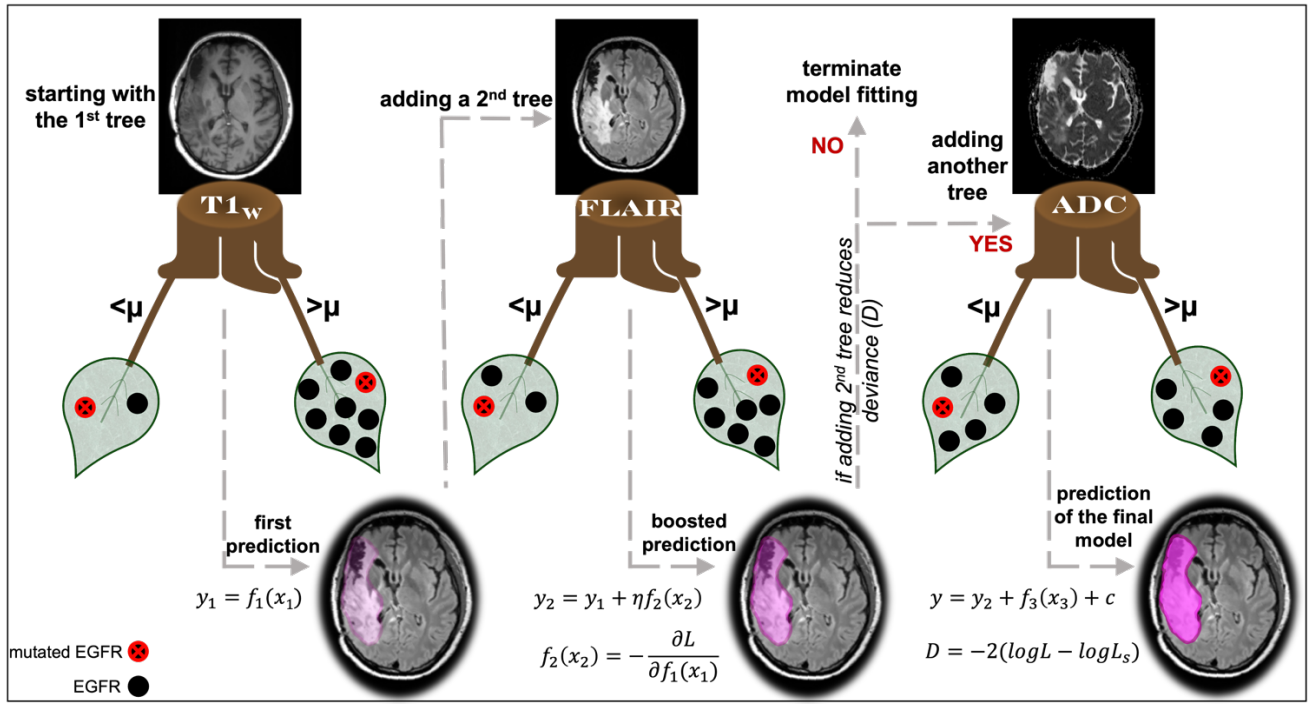

*Figure 2. Architectural diagram for gradient boosting algorithm of GAM. Predictors: $T1_w$, $T2_w$-FLAIR and ADC; Outcome: EGFR mutation status; μ: mean, $y_n$: observed response, $f_i(x_n)$: prediction after tree i for observation $x_n$, η: learning rate, L: lost function, and D: deviance. (GAM: generalized additive model; $T1_w$: T1-weighted; EGFR: epidermal growth factor receptor; FLAIR: fluid-attenuated inversion recovery; ADC: apparent diffusion coefficient)*

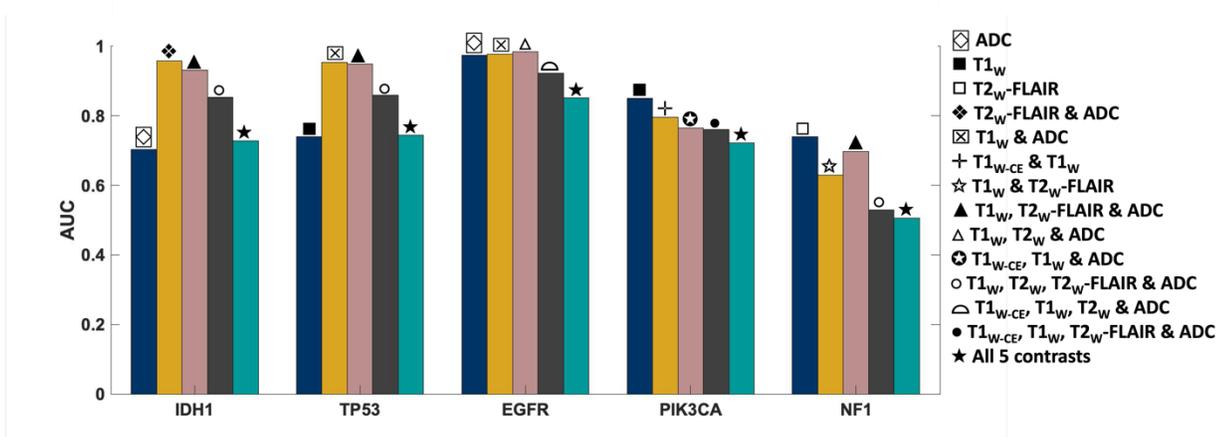

*Figure 3. AUC scores of the optimal MRI contrast combinations for predicting each mutation status. Navy represents single contrast, Mustard for double contrasts, Pink for triple contrasts, Smoke for quadruple contrasts, and Teal indicates quintuple contrasts.*
*(AUC: area under curve; MRI: magnetic resonance imaging; IDH1: isocitrate dehydrogenase 1; TP53: tumor protein 53; EGFR: epidermal growth factor receptor; PIK3CA: phosphatidylinositol-4,5-bisphosphate 3-kinase catalytic subunit alpha; NF1: neurofibromatosis type 1; ADC: apparent diffusion coefficient; $T1_W$: T1-weighted; $T2_W$-FLAIR: T2-weighted fluid attenuated inversion recovery; $T1_{W-CE}$: contrast enhanced T1-weighted; $T2_W$: T2-weighted).*

## Results

In this work, the ability of individual MR contrasts to predict genomic features in glioma was assessed by conducting ROC analyses. The observed average ACC values ranged between 0.71 and 0.83, indicative of statistical robustness. The AUC scores for each of the 31 combinations pertaining to each genetic target are detailed in the Supplementary Materials. Figure 3 shows the contrasts yielding the highest AUC scores and the changes in AUC scores upon adding more contrasts as predictors. Notably, $T1_W$ and ADC were the most frequently occurring contrasts within optimal combinations, with $T2_W$-FLAIR following closely. Every group of triple and quadruple contrasts included $T1_W$ and ADC, while double contrast groups had at least one of them. The ROC curves for these optimal combinations are plotted in Figure 4. The ROC figures demonstrate the model's outstanding performance when predicting IDH1, TP53, and EGFR, as the curves achieve a TPR (True Positive Rate) of 1 with a minimal FPR. Among all tested combinations of MR contrasts, the specific triple combination of $T1_W$, ADC, and $T2_W$-FLAIR stood out for its remarkable predictive prowess across all five mutational targets. Figure 5 shows the multi-class ROC curve representing this combination. For the gene targets IDH1, TP53, and EGFR, the AUC scores exceeded 0.9, signifying excellent predictivity. Additionally, the scores for PIK3CA and NF1 registered above 0.7, indicative of acceptable performance.

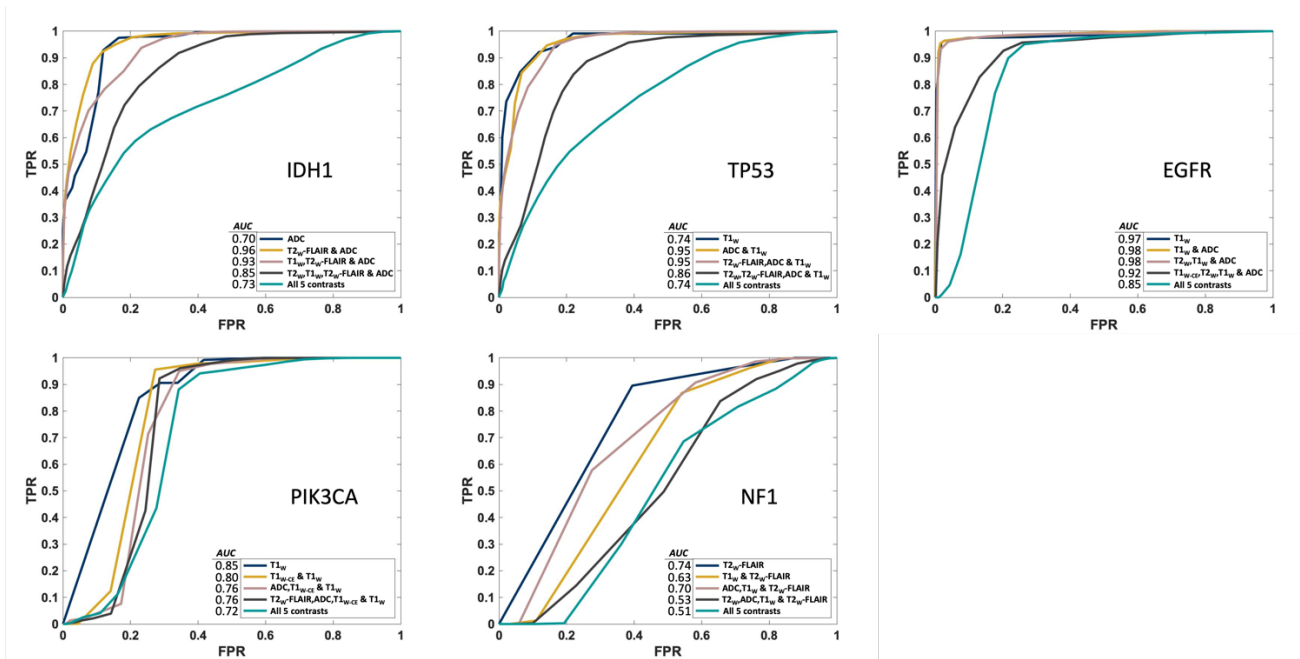

*Figure 4. ROC curves of the optimal MRI contrast combinations for predicting each mutation status. Navy represents single contrast, Mustard for double contrasts, Pink for triple contrasts, Smoke for quadruple contrasts, and Teal indicates quintuple contrasts. The AUC score for each curve is indicated next to the predictors' names in the legend for each mutational target.*
*(ROC: receiving operating characteristic; MRI: magnetic resonance imaging; AUC: area under curve; IDH1: isocitrate dehydrogenase 1; TP53: tumor protein 53; EGFR: epidermal growth factor receptor; PIK3CA: phosphatidylinositol-4,5-bisphosphate 3-kinase catalytic subunit alpha; NF1: neurofibromatosis type 1; ADC: apparent diffusion coefficient; $T1_W$: T1-weighted; $T2_W$-FLAIR: T2-weighted fluid attenuated inversion recovery; $T1_{W-CE}$: contrast enhanced T1-weighted; $T2_W$: T2-weighted; TPR: true positive rate; FPR: false positive rate).*

**Discussion**

In this study, a regression-based GAM was developed and tested utilizing clinical MR images to predict the mutational status of select glioma genomic targets. All combinations of $T1_W$, $T1_{W-CE}$, $T2_W$, $T2_W$-FLAIR, and ADC were encompassed in the model training as predictors, with the mutation status of IDH1, TP53, EGFR, PIK3CA, and NF1 as outcomes. The training was conducted over 10 patients using the leave-one-patient-out method. ROC curve analysis was completed for each of the 31 total combinations, and AUC scores were calculated. $T1_W$ and ADC emerged as the foremost contrasts, with $T2_W$-FLAIR ranking next in high predictive power for all five targets, resulting in AUC scores as high as 0.98, accompanied by robust ACC values.

The selected MRI sequences for this study are integral components of the clinical neuroimaging routine for glioma patients. Their incorporation into the clinical practice is not arbitrary but rather grounded in a robust body of literature that emphasizes their clinical relevance. For instance, high-resolution 3D $T1_W$ and $T2_W$ sequences, along with $T1_{W-CE}$, are essential for detecting abnormalities in the blood-brain barrier (BBB) and areas with increased vascularity.[25,26] The $T2_W$-FLAIR technique enhances the visibility of lesions in the periventricular and peripheral subcortical regions by suppressing the CSF signal and reducing the contrast difference between gray and white matter.[27] This allows for the differentiation of vasogenic edema from normal brain fluids and aids in identifying infiltrative microscopic pathology.[28] The apparent diffusion coefficient (ADC) map, derived from the DWI signal, provides insights into the glioma tumor cellularity,[29] and its utility in grading these tumors has been well studied.[30,31]

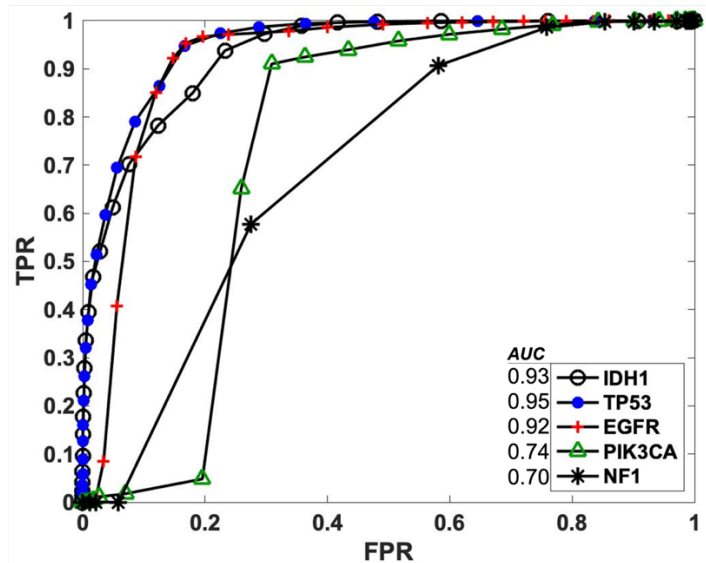

*Figure 5. Multi-class ROC curve of $T1_W$, $T2_W$-FLAIR and ADC as predictors for all 5 mutations. The AUC score for each curve is indicated next to the predictors' names in the legend for each mutational target.*
*(ROC: receiving operating characteristic; $T1_W$: T1-weighted; $T2_W$-FLAIR: T2-weighted fluid attenuated inversion recovery; ADC: apparent diffusion coefficient; AUC: area under curve; IDH1: isocitrate dehydrogenase 1; TP53: tumor protein 53; EGFR: epidermal growth factor receptor; PIK3CA: phosphatidylinositol-4,5-bisphosphate 3-kinase catalytic subunit alpha; NF1: neurofibromatosis type 1; TPR: true positive rate; FPR: false positive rate).*

As shown in Figures 3 and 4, the $T2_W$ signal did not emerge as a leading predictor for any genetic targets unless paired with other MR contrasts. Its most pronounced contribution was only seen in quadruple and quintuple contrast groups. This outcome is somewhat unexpected since $T2_W$ images are routinely used in clinical brain tumor diagnosis.[32] However, their primary application in such contexts is for qualitative visual assessment, not the signal quantification that was employed in building this model. One explanation for this outcome could be the substitution of $T2_W$ with Gadolinium-enhanced $T2_W$ ($T2_{W-CE}$) images for two subjects in the training. This change was implemented because two patients lacked $T2_W$ images in their most recent pre-surgical imaging dataset and were instead scanned with $T2_{W-CE}$ sequences. In addition, while Gadolinium-based contrast agents can diminish the $T2_W$ signal, a notable decrease[33] only occurs when the administered dose is higher than the FDA-approved limit (1 mmol/kg)[34,35] which was not the case in this study.

Gliomas exhibit a spectrum of genetic alterations that vary depending on their grade and specific subtype.[36] The prognostic implications of these genetic markers have been extensively studied.[37] For example, IDH1 mutations are commonly found in lower-grade gliomas and certain glioblastomas[38] whereas alterations in EGFR and mutations in TP53 span across various glioma grades.[39-41] In alignment with the existing literature, excellent AUC scores for IDH1, TP53, and EGFR were obtained in this study. The ROC curves for these three markers, using an optimal MR contrast combination, are presented in Figure 5.

Furthermore, mutations in both the PIK3CA and NF1 genes have been frequently associated with glioblastoma, glioma WHO grade IV.[42,43] These mutations are rare in lower-grade gliomas, with the exception of NF1 alterations in pediatric optic nerve gliomas.[44] The calculated AUC score using PIK3CA and NF1 as dependent variables was near 0.7, as shown in Figures 2-5. This comparatively weaker predictive power is consistent with the literature, considering that only half of our cohort had a glioblastoma diagnosis (as detailed in Table 1), and the data were not stratified by glioma grade during model construction.

**Conclusion**

This study contributes to the body of evidence integrating MRI imaging and stereotactic biopsy sampling with advanced statistical modeling to accurately predict glioma genomic targets. This type of experimental design is challenging to administer but provides robust and highly controlled data collection on intra-tumoral heterogeneity, the driving factor in human brain tumor treatment resistance and recurrence. The regression-based GAM presented here exhibits remarkable promise in using $T1_W$, ADC, and $T2_W$-FLAIR, especially for key mutations like IDH1, TP53, and EGFR. However, certain nuances, such as the predictability for markers such as PIK3CA and NF1, suggest the need for further refinement and larger cohorts for improved accuracy. This study lays a foundation for future work in non-invasive tumor diagnostics, enhancing treatment precision while minimizing patient risk. As the medical community gravitates towards individualized treatment plans, such innovative approaches will be instrumental in revolutionizing patient care in neuro-oncology.